\documentclass[italian,english]{article}
\usepackage[latin9]{inputenc}
\usepackage{color}
\usepackage{amsmath}
\usepackage{amssymb}

\newcommand{\lyxaddress}[1]{
\par {\raggedright #1
\vspace{1.4em}
\noindent\par}
}

\usepackage{babel}

\begin{document}

\title{\textbf{Interferometric detection of gravitational waves: the definitive
test for General Relativity}}

\author{\textbf{Christian Corda }}

\maketitle

\lyxaddress{\begin{center}
Associazione Scientifica Galileo Galilei, Via Pier Cironi 16 - 59100
PRATO, Italy 
\par\end{center}}

\begin{center}
\textit{E-mail address:} \textcolor{blue}{cordac.galilei@gmail.com} 
\par\end{center}
\begin{abstract}
Even if Einstein's General Relativity achieved a great success and
overcame lots of experimental tests, it also showed some shortcomings
and flaws which today advise theorists to ask if it is the definitive
theory of gravity. In this essay we show that, if advanced projects
on the detection of Gravitational Waves (GWs) will improve their sensitivity,
allowing to perform a GWs astronomy, accurate angular and frequency
dependent response functions of interferometers for GWs arising from
various Theories of Gravity, i.e. General Relativity and Extended
Theories of Gravity, will be the definitive test for General Relativity.
The papers which found this essay have been the world's most cited
in the official Astroparticle Publication Review of ASPERA during
the 2007 with 13 citations.\end{abstract}
\begin{itemize}
\item Essay dedicated to my wife Maria and written for the 2009 Gravity
Research Foundation Awards: Honorable Mention Winner
\end{itemize}
Recently, the data analysis of interferometric GWs detectors has been
started (for the current status of GWs interferometers see \cite{key-1})
and the scientific community aims in a first direct detection of GWs
in next years. 

Detectors for GWs will be important for a better knowledge of the
Universe and either to confirm or rule out the physical consistency
of General Relativity or of any other theory of gravitation \cite{key-2,key-3,key-4,key-5,key-6,key-7}.
In fact, in the context of Extended Theories of Gravity, some differences
between General Relativity and the others theories can be pointed
out starting by the linearized theory of gravity \cite{key-2,key-3,key-4,key-5,key-6,key-7,key-8,key-9,key-10,key-11,key-12}.
In this picture, detectors for GWs are in principle sensitive also
to a hypothetical \textit{scalar} component of gravitational radiation,
that appears in extended theories of gravity like scalar-tensor gravity
\cite{key-4,key-8,key-9,key-10,key-12}, bi-metric theory \cite{key-5},
high order theories \cite{key-2,key-3,key-6,key-7}, Brans-Dicke theory
\cite{key-13} and string theory \cite{key-14}.

Motivations of extending General Relativity arise from the fact that,
even if Einstein's Theory \cite{key-15} achieved a great success
(see for example the opinion of Landau who says that General Relativity
is, together with Quantum Field Theory, the best scientific theory
of all \cite{key-16}) and overcame lots of experimental tests \cite{key-15},
it also showed some shortcomings and flaws which today advise theorists
to ask if it is the definitive theory of gravity \cite{key-17,key-18}.
Differently from other field theories like the electromagnetic theory,
General Relativity is very difficult to be quantized. This fact rules
out the possibility of treating gravitation like other quantum theories,
and precludes the unification of gravity with other interactions.
At the present time, it is not possible to realize a consistent Quantum
Gravity Theory which leads to the unification of gravitation with
the other forces \cite{key-17,key-18}. 

On the other hand, one can define \textit{Extended Theories of Gravity}
those semiclassical theories where the Lagrangian is modified, in
respect to the standard Einstein-Hilbert gravitational Lagrangian,
adding high-order terms in the curvature invariants (terms like $R^{2}$,
$R^{\alpha\beta}R_{\alpha\beta}$, $R^{\alpha\beta\gamma\delta}R_{\alpha\beta\gamma\delta}$,
$R\Box R$, $R\Box^{k}R$) or terms with scalar fields non minimally
coupled to geometry (terms like $\phi^{2}R$) \cite{key-17,key-18}.
In general, one has to emphasize that terms like those are present
in all the approaches to perform the unification between gravity and
other interactions. More, from a cosmological point of view, such
modifies of General Relativity generate inflationary frameworks which
are very important as they solve lots of problems of the Standard
Universe Model \cite{key-19}. Note that we are not telling that General
Relativity is wrong. It is well known that, even in the context of
Extended Theories, General Relativity remains the most important part
of the structure \cite{key-4,key-7,key-17,key-18}. We are only trying
to understand if weak modifies on such a structure could be needed
to solve some theoretical and observing problems \cite{key-17,key-18}.
In this picture, we also recall that even Einstein told that General
Relativity could not be definitive \cite{key-28}. In fact, during
his famous research on the Unified Field Theory, he tried to realize
a theory that he called {}``Generalized Theory of Gravitation'',
and he said that  mathematical difficulties precluded him to obtain
the final equations \cite{key-28}.

In the general context of cosmological evidences, there are also other
considerations which suggest an extension of General Relativity. As
a matter of fact, the accelerated expansion of the Universe, which
is today observed, shows that cosmological dynamic is dominated by
the so called Dark Energy, which gives a large negative pressure.
This is the standard picture, in which such new ingredient is considered
as a source of the \textit{right side} of the field equations. It
should be some form of un-clustered non-zero vacuum energy which,
together with the clustered Dark Matter, drives the global dynamics.
This is the so called {}``concordance model'' ($\Lambda$CDM) which
gives, in agreement with the CMBR, LSS and SNeIa data, a good tapestry
of the today observed Universe, but presents several shortcomings
as the well known {}``coincidence'' and {}``cosmological constant''
problems \cite{key-20}. An alternative approach is changing the \textit{left
side} of the field equations, seeing if observed cosmic dynamics can
be achieved extending General Relativity \cite{key-17,key-18,key-21,key-22}.
In this different context, it is not required to find out candidates
for Dark Energy and Dark Matter, that, till now, have not been found,
but only the {}``observed'' ingredients, which are curvature and
baryon matter, have to be taken into account. Considering this point
of view, one can think that gravity is different at various scales
\cite{key-21} and a room for alternative theories is present. In
principle, the most popular Dark Energy and Dark Matter models can
be achieved considering $f(R)$ theories of gravity, where $R$ is
the Ricci curvature scalar, and/or Scalar-Tensor Gravity \cite{key-17,key-18,key-22}.

In this essay we show that, if advanced projects on the detection
of GWs will improve their sensitivity, allowing to perform a GWs astronomy
\cite{key-1}, accurate angular and frequency dependent response functions
of interferometers for GWs arising from various Theories of Gravity,
i.e. General Relativity and Extended Theories of Gravity, will be
the definitive test for General Relativity \cite{key-2,key-3,key-4,key-5,key-6,key-7,key-10,key-11,key-12}.
The papers which found this essay have been the world's most cited
in the official Astroparticle Publication Review of ASPERA during
the 2007 with 13 citations \cite{key-23}. We recall that ASPERA is
the network of national government agencies responsible for coordinating
and funding national research efforts in Astroparticle Physics, see
\cite{key-30}.

Working with $G=1$, $c=1$ and $\hbar=1$ (natural units), the line
element for a GW arising from standard General Relativity and propagating
in the $z$ direction is \cite{key-1,key-15,key-24,key-25} 

\begin{equation}
ds^{2}=dt^{2}-dz^{2}-(1+h_{+})dx^{2}-(1-h_{+})dy^{2}-2h_{\times}dxdy,\label{eq: metrica TT totale}\end{equation}

where $h_{+}(t+z)$ and $h_{\times}(t+z)$ are the weak perturbations
due to the $+$ and the $\times$ polarizations which are expressed
in terms of synchronous coordinates in the Transverse Traceless (TT)
gauge \cite{key-15}. In \cite{key-24,key-25} it has been shown that
the total frequency and angular dependent response function (i.e.
the detector pattern) to the $+$ polarization of an interferometer
with arms in the $u$ and $v$ directions in respect to the propagating
GW is:\begin{align}
\tilde{H}^{+}(\omega) & \equiv\Upsilon_{u}^{+}(\omega)-\Upsilon_{v}^{+}(\omega)\nonumber \\
 & =\frac{(\cos^{2}\theta\cos^{2}\phi-\sin^{2}\phi)}{2L}\tilde{H}_{u}(\omega,\theta,\phi)-\frac{(\cos^{2}\theta\sin^{2}\phi-\cos^{2}\phi)}{2L}\tilde{H}_{v}(\omega,\theta,\phi),\label{eq: risposta totale Virgo +}\end{align}

that, in the low frequencies limit ($\omega\rightarrow0$) gives the
well known low frequency response function of \cite{key-26,key-27}
for the $+$ polarization: 

\begin{equation}
\tilde{H}^{+}(\omega)=\frac{1}{2}(1+\cos^{2}\theta)\cos2\phi+O\left(\omega\right)\,.\label{eq: risposta totale approssimata}\end{equation}

For a sake of clearness, let us sketch the derivation of eq. (\ref{eq: risposta totale Virgo +}).

The rotation in respect to the $u$ and $v$ directions is

\begin{equation}
\begin{array}{ccc}
x & = & -u\cos\theta\cos\phi-v\cos\theta\sin\phi+w\sin\theta\\
\\y & = & u\sin\phi-v\cos\phi\\
\\z & = & u\sin\theta\cos\phi+v\sin\theta\sin\phi+w\cos\theta,\end{array}\label{eq: rotazione 2}\end{equation}

with the correspondent line element transformation \cite{key-24,key-25}

\begin{equation}
g^{ik}=\frac{\partial x^{i}}{\partial x'^{l}}\frac{\partial x^{k}}{\partial x'^{m}}g'^{lm},\label{eq: trasformazione metrica}\end{equation}

By using (\ref{eq: rotazione 2}) and (\ref{eq: trasformazione metrica}),
and taking into account only the $+$ polarization, the line element
in the $\overrightarrow{u}$ direction becomes:

\begin{equation}
ds^{2}=-dt^{2}+[1+(\cos^{2}\theta\cos^{2}\phi-\sin^{2}\phi)h_{+}(t+u\sin\theta\cos\phi)]du^{2}.\label{eq: metrica + lungo u}\end{equation}

Following \cite{key-24,key-25,key-29}, a good way to analyse variations
in the proper distance (time) is by means of {}``bouncing photons''.
A photon can be launched from the interferometer's beam-splitter to
be bounced back by the mirror. The {}``bouncing photons analysis''
was created by Rakhmanov in \cite{key-29}. Actually, it has strongly
generalized to angular dependences, scalar waves and massive GWs in
\cite{key-2,key-4,key-12,key-24,key-25}.

The condition for null geodesics ($ds^{2}=0$) in eq. (\ref{eq: metrica + lungo u})
gives the coordinate velocity of the photon:

\begin{equation}
v_{p}^{2}\equiv(\frac{du}{dt})^{2}=\frac{1}{[1+(\cos^{2}\theta\cos^{2}\phi-\sin^{2}\phi)h_{+}(t+u\sin\theta\cos\phi)]},\label{eq: velocita' fotone u}\end{equation}

which will be used for calculations of the photon propagation time
between the beam-splitter and the mirror \cite{key-24,key-25,key-29}.
We assume that the beam splitter is located in the origin of the new
coordinate system (i.e. $u_{b}=0$, $v_{b}=0$, $w_{b}=0$). Being
in the TT gauge, the coordinates of the beam-splitter $u_{b}=0$ and
of the mirror $u_{m}=L$ do not change under the influence of the
GW\cite{key-15,key-25}, thus the duration of the forward trip can
be written as

\begin{equation}
T_{1}(t)=\int_{0}^{L}\frac{du}{v_{p}(t'+u\sin\theta\cos\phi)},\label{eq: durata volo}\end{equation}

with 

\begin{center}
$t'=t-(L-u)$.
\par\end{center}

In the last equation $t'$ is the delay time (i.e. $t$ is the time
at which the photon arrives in the position $L$, so $L-u=t-t'$).

At first order in $h_{+}$ this integral can be approximated with

\begin{equation}
T_{1}(t)=T+\frac{\cos^{2}\theta\cos^{2}\phi-\sin^{2}\phi}{2}\int_{0}^{L}h_{+}(t'+u\sin\theta\cos\phi)du,\label{eq: durata volo andata approssimata u}\end{equation}

where, as we are using natural units, $T=L$ is the transit time of
the photon in absence of the GW. Similarly, the duration of the return
trip will be\begin{equation}
T_{2}(t)=T+\frac{\cos^{2}\theta\cos^{2}\phi-\sin^{2}\phi}{2}\int_{L}^{0}h_{+}(t'+u\sin\theta\cos\phi)(-du),\label{eq: durata volo ritorno approssimata u}\end{equation}

though now the delay time is 

\begin{center}
$t'=t-(u-l)$.
\par\end{center}

The round-trip time will be the sum of $T_{2}(t)$ and $T_{1}[t-T_{2}(t)]$.
The latter can be approximated by $T_{1}(t-T)$ because the difference
between the exact and the approximate values is second order in $h_{+}$.
Then, to first order in $h_{+}$, the duration of the round-trip will
be

\begin{equation}
T_{r.t.}(t)=T_{1}(t-T)+T_{2}(t).\label{eq: durata round trip}\end{equation}

By using eqs. (\ref{eq: durata volo andata approssimata u}) and (\ref{eq: durata volo ritorno approssimata u})
one sees immediately that deviations of this round-trip time (i.e.
proper distance) from its unperturbed value are given by

\begin{equation}
\begin{array}{c}
\delta T(t)=\frac{\cos^{2}\theta\cos^{2}\phi-\sin^{2}\phi}{2}\int_{0}^{L}[h_{+}(t-2T-u(1-\sin\theta\cos\phi))+\\
\\+h_{+}(t+u(1+\sin\theta\cos\phi))]du.\end{array}\label{eq: variazione temporale in u}\end{equation}

Now, using the Fourier transform of the $+$ polarization of the field,
defined by

\begin{equation}
\tilde{h}_{+}(\omega)\equiv\int_{-\infty}^{\infty}dth_{+}(t)\exp(i\omega t),\label{eq: TF}\end{equation}

one obtains in the frequency domain:

\begin{equation}
\delta\tilde{T}(\omega)=\frac{1}{2}(\cos^{2}\theta\cos^{2}\phi-\sin^{2}\phi)\tilde{H}_{u}(\omega,\theta,\phi)\tilde{h}_{+}(\omega),\label{eq: segnale in frequenza lungo u}\end{equation}

where

\begin{equation}
\begin{array}{c}
\tilde{H}_{u}(\omega,\theta,\phi)=\frac{-1+\exp(2i\omega L)}{2i\omega(1+\sin^{2}\theta\cos^{2}\phi)}+\\
\\+\frac{-\sin\theta\cos\phi((1+\exp(2i\omega L)-2\exp i\omega L(1-\sin\theta\cos\phi)))}{2i\omega(1+\sin\theta\cos^{2}\phi)}\end{array}\label{eq: fefinizione Hu}\end{equation}

and we immediately see that $\tilde{H}_{u}(\omega,\theta,\phi)\rightarrow L$
when $\omega\rightarrow0$.

Thus, by defining the {}``signal'' in the $u$ arm like $S(\omega)\equiv\frac{\delta\tilde{T}(\omega)}{2T},$
the total response function of this arm of the interferometer to the
$+$ component is:

\begin{equation}
\Upsilon_{u}^{+}(\omega)\equiv\frac{S(\omega)}{\tilde{h}_{+}(\omega)}=\frac{(\cos^{2}\theta\cos^{2}\phi-\sin^{2}\phi)}{2L}\tilde{H}_{u}(\omega,\theta,\phi).\label{eq: risposta + lungo u}\end{equation}

In the same way, one gets the response function of the $v$ arm of
the interferometer to the $+$ polarization: 

\begin{equation}
\Upsilon_{v}^{+}(\omega)=\frac{(\cos^{2}\theta\sin^{2}\phi-\cos^{2}\phi)}{2L}\tilde{H}_{v}(\omega,\theta,\phi)\label{eq: risposta + lungo v}\end{equation}

where, now 

\begin{equation}
\begin{array}{c}
\tilde{H}_{v}(\omega,\theta,\phi)=\frac{-1+\exp(2i\omega L)}{2i\omega(1+\sin^{2}\theta\sin^{2}\phi)}+\\
\\+\frac{-\sin\theta\sin\phi((1+\exp(2i\omega L)-2\exp i\omega L(1-\sin\theta\sin\phi)))}{2i\omega(1+\sin^{2}\theta\sin^{2}\phi)},\end{array}\label{eq: fefinizione Hv}\end{equation}

with $\tilde{H}_{v}(\omega,\theta,\phi)\rightarrow L$ when $\omega\rightarrow0$. 

The total response function is defined like the difference between
(\ref{eq: risposta + lungo u}) and (\ref{eq: risposta + lungo v}),
thus one obtains immediately eq. (\ref{eq: risposta totale Virgo +}).

The same analysis works for the $\times$ polarization (see \cite{key-24,key-25}
for details). One obtains that the total frequency and angular dependent
response function of an interferometer to the $\times$ polarization
is:

\begin{equation}
\tilde{H}^{\times}(\omega)=\frac{-\cos\theta\cos\phi\sin\phi}{L}[\tilde{H}_{u}(\omega,\theta,\phi)+\tilde{H}_{v}(\omega,\theta,\phi)],\label{eq: risposta totale Virgo per}\end{equation}
that, in the low frequencies limit ($\omega\rightarrow0$), gives
the low frequency response function of \cite{key-26,key-27} for the
$\times$ polarization: \begin{equation}
\tilde{H}^{\times}(\omega)=-\cos\theta\sin2\phi+O\left(\omega\right)\,.\label{eq: risposta totale approssimata 2}\end{equation}

The case of massless Scalar-Tensor Gravity has been discussed in \cite{key-4,key-12}
with a {}``bouncing photons analysis'' similar to the previous one
. In this case, the line-element in the TT gauge can be extended with
one more polarization, labelled with $\Phi(t+z)$, i.e.

\begin{equation}
ds^{2}=dt^{2}-dz^{2}-(1+h_{+}+\Phi)dx^{2}-(1-h_{+}+\Phi)dy^{2}-2h_{\times}dxdy.\label{eq: metrica TT super totale}\end{equation}

The total frequency and angular dependent response function of an
interferometer to this {}``scalar'' polarization is \cite{key-4,key-12}\begin{align}
\tilde{H}^{\Phi}(\omega) & =\frac{\sin\theta}{2i\omega L}\{\cos\phi[1+\exp(2i\omega L)-2\exp i\omega L(1+\sin\theta\cos\phi)]+\nonumber \\
 & -\sin\phi[1+\exp(2i\omega L)-2\exp i\omega L(1+\sin\theta\sin\phi)]\}\,,\label{eq: risposta totale Virgo scalar}\end{align}

that, in the low frequencies limit ($\omega\rightarrow0$), gives
the low frequency response function of \cite{key-9,key-14} for the
$\Phi$ polarization: \textbf{\begin{equation}
\tilde{H}^{\Phi}(\omega)=-\sin^{2}\theta\cos2\phi+O(\omega).\label{eq: risposta totale approssimata scalar}\end{equation}
}

In \cite{key-2,key-3,key-4,key-7} it has also been shown that, in
the framework of GWs, the cases of massive Scalar-Tensor Gravity and
$f(R)$ theories are totally equivalent (this is not surprising as
it is well known that there is a more general conformal equivalence
between Scalar-Tensor Gravity and $f(R)$ theories, even if there
is a large debate on the possibility that such a conformal equivalence
should be a \emph{physical} equivalence too \cite{key-17,key-18,key-21}).
In such cases, because of the presence of a small mass, a longitudinal
component is present in the third polarization, thus it is impossible
to extend the TT gauge to the third mode \cite{key-2,key-3,key-4,key-6,key-7}.
But, by using gauge transformations, one can put the line-element
due to such a third scalar mode in a conformally flat form \cite{key-2,key-3,key-4,key-6,key-7}: 

\begin{equation}
ds^{2}=[1+\Phi(t-v_{G}z)](-dt^{2}+dz^{2}+dx^{2}+dy^{2}).\label{eq: metrica puramente scalare}\end{equation}

If the interferometer arm is parallel to the propagating GW, the longitudinal
response function, which has been obtained in \cite{key-2,key-7}
with the {}``bouncing photons analysis'', associated to such a massive
mode is \begin{equation}
\begin{array}{c}
\Upsilon_{l}(\omega)=\frac{1}{m^{4}\omega^{2}L}(\frac{1}{2}(1+\exp[2i\omega L])m^{2}\omega^{2}L(m^{2}-2\omega^{2})+\\
\\-i\exp[2i\omega L]\omega^{2}\sqrt{-m^{2}+\omega^{2}}(4\omega^{2}+m^{2}(-1-iL\omega))+\\
\\+\omega^{2}\sqrt{-m^{2}+\omega^{2}}(-4i\omega^{2}+m^{2}(i+\omega L))+\\
\\+\exp[iL(\omega+\sqrt{-m^{2}+\omega^{2}})](m^{6}L+m^{4}\omega^{2}L+8i\omega^{4}\sqrt{-m^{2}+\omega^{2}}+\\
\\+m^{2}(-2L\omega^{4}-2i\omega^{2}\sqrt{-m^{2}+\omega^{2}}))+2\exp[i\omega L]\omega^{3}(-3m^{2}+4\omega^{2})\sin[\omega L]),\end{array}\label{eq: risposta totale lungo z massa}\end{equation}

where $m$ in eq. (\ref{eq: risposta totale lungo z massa}) is the
small mass of the particle associated to the GW and $v_{G}$ in eq.
(\ref{eq: metrica puramente scalare}) is the particle's velocity
(the group velocity in terms of a wave-packet \cite{key-2,key-7}).
The relation mass-velocity is $m=\sqrt{(1-v_{G}^{2})}\omega,$ see
\cite{key-2,key-7} for details.

Thus, if advanced projects on the detection of GWs will improve their
sensitivity allowing to perform a GWs astronomy (this is due because
signals from GWs are quite weak) \cite{key-1}, one will only have
to look the interferometer response functions to understand if General
Relativity is the definitive theory of gravity. In fact, if only the
two response functions (\ref{eq: risposta totale Virgo +}) and (\ref{eq: risposta totale Virgo per})
will be present, we will conclude that General Relativity is definitive.
If the response function (\ref{eq: risposta totale Virgo scalar})
will be present too, we will conclude that massless Scalar - Tensor
Gravity is the correct theory of gravitation. Finally, if a longitudinal
response function will be present, i.e. Eq. (\ref{eq: risposta totale lungo z massa})
for a wave propagating parallel to one interferometer arm, or its
generalization to angular dependences, we will learn that the correct
theory of gravity will be massive Scalar - Tensor Gravity which is
equivalent to $f(R)$ theories. In any case, such response functions
will represent the definitive test for General Relativity. This is
because General Relativity is the only gravity theory which admits
only the two response functions (\ref{eq: risposta totale Virgo +})
and (\ref{eq: risposta totale Virgo per}) \cite{key-4,key-7,key-17,key-18}.
Such response functions correspond to the two {}``canonical'' polarizations
$h_{+}$ and $h_{\times}.$ Thus, if a third polarization will be
present, a third response function will be detected by GWs interferometers
and this fact will rule out General Relativity like the definitive
theory of gravity. 

Resuming, in this essay we have shown that, by assuming that advanced
projects on the detection of GWs will improve their sensitivity allowing
to perform a GWs astronomy, accurate angular and frequency dependent
response functions of interferometers for gravitational waves arising
from various Theories of Gravity, i.e. General Relativity and Extended
Theories of Gravity, will be the definitive test for General Relativity.

\subsubsection*{Acknowledgements}

I thank my wife Maria for her love and her support in my research
work. This essay is dedicated to her. The Associazione Scientifica
Galileo Galilei has to be thanked for supporting this essay.


\begin{thebibliography}{30}
\bibitem{key-1}A. Giazotto - Journ. of Phys., Conf. Series 120, 032002
(2008) 

\bibitem[2]{key-2}C. Corda - J. Cosmol. Astropart. Phys. JCAP04009
(2007)

\bibitem[3]{key-3}\foreignlanguage{italian}{C. Corda - Int. Journ.}
Mod. Phys. A 23, 10, 1521-1535 (2008)

\bibitem[4]{key-4}Capozziello S and C. Corda - Int. J. Mod. Phys.
D \textbf{15,} 1119 -1150 (2006) 

\bibitem[5]{key-5}C. Corda - Astropart. Phys. 28, 2, 247-250 (2007)

\bibitem[6]{key-6}C. Corda - Astropart. Phys. 30, 4 209-215 (2008)

\bibitem[7]{key-7}S. Capozziello, C. Corda and M. F. De Laurentis
- Phys.Lett. B, 669, 5, 255-259, (2008) 

\bibitem[8]{key-8}T. Damour and G. Esposito-Farese - Class. Quant.
Grav. \textbf{9} 2093-2176 (1992);  

\bibitem[9]{key-9}\foreignlanguage{italian}{C. Corda - Int. Journ.}
Mod. Phys. A 22, 26, 4859-4881 (2007)

\bibitem[10]{key-10}S. Capozziello, C. Corda and M. F. De Laurentis
- Mod. Phys. Lett. A 22, 35, 2647-2655 (2007)

\bibitem[11]{key-11}S. Capozziello, C. Corda and M. F. De Laurentis
- Mod. Phys. Lett. A 22, 15, 1097-1104 (2007)

\bibitem[12]{key-12}C. Corda - Mod. Phys. Lett. A No. 22, 23, 1727-1735
(2007)

\bibitem[13]{key-13}C. Brans and R. H. Dicke - Phys. Rev. 124, 925
(1961)

\bibitem[14]{key-14}N. Bonasia and M. Gasperini - Phys. Rew. D \textbf{71}
104020 (2005)

\bibitem[15]{key-15}\foreignlanguage{italian}{C. W. Misner, K. S.
Thorne and J. A. Wheeler - {}``Gravitation'' - W.H.Feeman and Company
- 1973} 

\bibitem[16]{key-16}L. Landau and E. Lifsits - {}``Teoria dei campi''
- Editori riuniti edition III (1999) 

\bibitem[17]{key-17}S. Capozziello and M. Francaviglia - Gen. Rel.
Grav. 40, 2-3, 357 (2008)

\bibitem[18]{key-18}V. Faraoni and T. P. Sotiriou - arXiv:0805.1249,
to appear in Rev. Mod. Phys.

\bibitem[19]{key-19}A. Starobinsky - Phys. Lett. B, 91, 99-102 (1980)

\selectlanguage{italian}%
\bibitem[20]{key-20}\foreignlanguage{english}{P. J. E. Peebles and
B. Ratra - Rev. Mod. Phys. 75 8559 (2003) }

\selectlanguage{english}%
\bibitem[21]{key-21}C. M. - Will \textit{Theory and Experiments in
Gravitational Physics}, Cambridge Univ. Press Cambridge (1993)

\bibitem[22]{key-22}T. Inagaky, S. Nojiri and S. D. Odintsov - J.
Cosmol. Astropart. Phys. JCAP0506010 (2005)

\bibitem[23]{key-23}http://www.aspera-eu.org/index.php?option=com\_content\&task=view\&id=125\&Itemid=99

\bibitem[24]{key-24}C. Corda - Astropart. Phys. \textbf{27,} No 6,
539-549 (2007)

\bibitem[25]{key-25}C. Corda - Int. J. Mod. Phys. D \textbf{16,}
9, 1497-1517  (2007)

\bibitem[26]{key-26}K. S. Thorne- \textit{300 Years of Gravitation}
- Ed. Hawking SW and Israel W Cambridge University Press p. 330 (1987)

\bibitem[27]{key-27}P. Saulson - \textit{Fundamental of Interferometric
Gravitational Waves Detectors} - World Scientific, Singapore (1994) 

\bibitem[28]{key-28} A. Pais - \textit{Subtle is the Lord} - Oxford
University Press (2005)

\bibitem[29]{key-29}Rakhmanov M - Phys. Rev. D \textbf{71} 084003
(2005) 

\bibitem[30]{key-30}http://www.aspera-eu.org
\end{thebibliography}
\end{document}